\documentclass[submit]{eps}
\usepackage{url}
\usepackage{hyperref}
\usepackage{multirow}
\usepackage{gensymb}
\usepackage{amsmath}  
\usepackage{amssymb}  
\usepackage{amsfonts} 
\usepackage{newtxtext,newtxmath}
\usepackage{latexsym}
\usepackage{graphicx}
\usepackage{lineno}
\title{Investigating the Reliability of the AfriTEC Model During the Descending Phase of Solar Cycle 24 Across East Africa}

\author{
	Efrem Amanuel Data, 1, Department of Physics, Wolaita Sodo University, Wolaita Sodo, Ethiopia\\
	2, Department of Geology, Wolaita Sodo University, Wolaita Sodo, Ethiopia\\
Email: \url{immanuelefawu@gmail.com} , ORCID: \url{0009-0006-3668-5445}	\\
Emmanuel Daudi Sulungu, Department of Physics, The University of Dodoma, Dodoma, Tanzania. \\Email: \url{edsulungu@gmail.com}, ORCID: \url{0000-0002-3234-0953}  \\
Daniel Izuikedinachi Okoh, 1, Department of Astronomy and Space Science, Technical University of Kenya, Nairobi, Kenya.\\
2, National Space Research and Development Agency (NASRDA), Abuja, Nigeria.\\ 
Email: \url{okodan2003@gmail.com}, ORCID: \url{0000-0001-8816-092X}	\\
Dejene Ambisa Terefe, Department of Space and Planetary Science (SPS), Space Science and Geospatial Institute (SSGI), Addis Ababa, Ethiopia.\\ 
Email: \url{dejeneambisa1@gmail.com}, ORCID: \url{0000-0002-5976-270X}
}

\abstract{This study investigates the reliability of the African Regional Ionospheric Total Electron Content (AfriTEC) model during the descending phase of Solar Cycle 24 (2016-2017) across East Africa. Using GNSS-derived TEC data from five equatorial and low-latitude stations MOIU, MAL2, ZAMB, ADIS, and MBAR the model's performance is assessed through statistical metrics, including Mean Absolute Error (MAE) and correlation coefficient ($r$). Results indicate that the AfriTEC model effectively captures the diurnal and seasonal behavior of TEC, particularly during equinoxes, with MAE values generally below 1.5 TECU and correlation coefficients exceeding 0.80. However, discrepancies emerge during solstice periods and post-sunset hours, reflecting the model's limitations in representing complex ionospheric processes such as the Equatorial Ionization Anomaly (EIA). To benchmark its performance, AfriTEC is also compared against the widely used NeQuick model. AfriTEC demonstrates superior regional adaptability and reduced error under most conditions, though it remains sensitive to localized ionospheric disturbances. These findings suggest that while AfriTEC is a valuable tool for ionospheric modeling in whole Africa especially at East African sector, enhancements incorporating real-time solar and geomagnetic indices could further improve its predictive capabilities.}
\begin{document}

\maketitle

\section{Introduction}

\noindent
The ionosphere is very important in space weather and radio signal propagation, making its study essential for improving the performance of Global Navigation Satellite Systems (GNSS) and related technologies. One of the key parameters that characterize the ionosphere is the Total Electron Content (TEC), which refers to the total number of electrons along a path between a GNSS satellite and a receiver. Accurate TEC modeling is important for mitigating ionospheric errors in GNSS applications, especially in equatorial and low-latitude regions where ionospheric variability is high \cite{aarons1993, komjathy1997}. East Africa, located around the geomagnetic equator, is particularly susceptible to ionospheric disturbances due to equatorial electrodynamics. The region experiences significant TEC fluctuations caused by solar activity, geomagnetic storms, and ionospheric irregularities, including EPBs. These variations can degrade the performance of satellite navigation systems and communication links \cite{aarons1993}.

\noindent
The years 2016-2017, part of the descending phase of Solar Cycle 24, were characterized by moderate to low levels of solar activity, with declining solar flux, sunspot numbers, and geomagnetic disturbances \cite{Pesnell2012}. This phase offers an ideal context for assessing the effectiveness of ionospheric models during relatively calm space weather conditions. Such evaluation is especially critical for the East African region, where ionospheric variability is highly sensitive to changes in solar and geomagnetic activity \cite{Olwendo2016}. The influence of solar activity often categorized by the phases of the solar cycle plays a significant role in shaping ionospheric behavior. The descending phase of Solar Cycle 24 (from 2015-2019) experienced weaker solar emissions than its peak and ascending phases, leading to distinct impacts on ionospheric dynamics across different latitudes \cite{clette2016}. Despite the overall decline in solar ionizing radiation during this period, particularly in 2016 and 2017, the ionosphere remained highly dynamic in equatorial and low-latitude zones like East Africa. This ongoing variability underscores the need for persistent TEC monitoring and advanced modeling strategies capable of capturing ionospheric behavior under all solar conditions \cite{olwendo2018}.

\noindent
 One of the effective ways to study ionospheric TEC is through GNSS-based observations. GNSS signals experience phase delays as they pass through the ionosphere, which can be exploited to calculate TEC. Several ground-based GNSS receivers have been deployed across East Africa to monitor these variations in TEC and improve the understanding of ionospheric behavior in this region \cite{yizengaw2014}. To model ionospheric TEC, various empirical and physics-based models have been developed at world. One such model is the AfriTEC model, designed to predict TEC over the African continent using GNSS data. AfriTEC helps fill observational gaps in Africa, where sparse data coverage has historically limited accurate TEC modeling. The performance of ionospheric models like AfriTEC during different solar and geomagnetic conditions is important for their validation and application. In East Africa, the AfriTEC model has shown promising results in representing TEC patterns during both quiet and disturbed conditions. However, the model's accuracy tends to vary with latitude, time of day, and levels of geomagnetic activity \cite{E1}.

\noindent
 Studies have shown that during the years 2016 and 2017, TEC levels in East Africa were generally lower than those observed during the solar maximum period, due to the decline in solar EUV radiation. Nonetheless, occurrences of nighttime EPBs and scintillation were still recorded, indicating the persistence of equatorial irregularities even during lower solar activity \cite{olwendo2018}. Modeling efforts during this period also benefited from the introduction of calibration techniques that reduced biases in GNSS-derived TEC values \cite{E3}. These calibrations are essential for improving the reliability of ionospheric models in East Africa \cite{yizengaw2014}. The Sunspot Number (SSN) and solar flux indices were significantly lower in 2016 and 2017, reflecting the descending nature of Solar Cycle 24. Despite this, ionospheric dynamics remained active due to thermospheric wind patterns and equatorial electrodynamics, underlining the importance of regional models over global ones for improved prediction in East Africa \cite{clette2016}.
\subsection{Statement of the Problem}
Accurate modeling of ionospheric TEC is essential for ensuring the reliability and precision of GNSS. However, East Africa remains one of the least investigated regions in global ionospheric studies, largely due to the scarcity of ground based GNSS infrastructure and historically limited observational data \cite{E1,Okoh19}. The region's location near the magnetic equator introduces highly dynamic electrodynamic processes and substantial TEC variability, especially during periods of low solar activity, such as the descending phase of Solar Cycle 24. These conditions present major challenges to both global and regional ionospheric models. Although global models like NeQuick are widely utilized, they often struggle to accurately capture localized ionospheric structures and temporal TEC fluctuations specific to the East African sector. This is increased positioning errors and signal degradation in GNSS-based applications. To overcome these limitations, regional models such as AfriTEC have been developed, leveraging locally sourced data for improved adaptability. However, their reliability under varying solar and geomagnetic conditions particularly during quiet periods remains insufficiently validated.
During 2016 and 2017, which represent the descending phase of Solar Cycle 24, the ionosphere over East Africa exhibited relatively low TEC values while continuing to experience equatorial plasma irregularities and scintillation phenomena. These persistent disturbances complicate TEC forecasting and necessitate a closer evaluation of model performance under such conditions. Consequently, this study aims to assess the effectiveness of the AfriTEC model across East Africa by comparing its predictions with those of the established NeQuick model. Special attention is given to AfriTEC's ability to capture diurnal and seasonal TEC variations during this low solar activity phase. Such an investigation is vital for enhancing TEC prediction capabilities, reducing GNSS positioning errors, and advancing regional space weather mitigation strategies tailored to East Africa's unique ionospheric environment.

\section{Data and Method}
In this paper work, the two types of ionospheric data were used the observed data, which is the GNSS, and AfriTEC model predicted ionospheric data from an East Africa IGS station during the descending phase of Solar Cycle 24 (2016-2017). The geomagnetic indices data containing F10.7, planetary Kp index, and Dst index data were also used within this study.
\subsection{GNSS Data}
\begin{figure}[h!]
		\centering
		\includegraphics[scale=0.85]{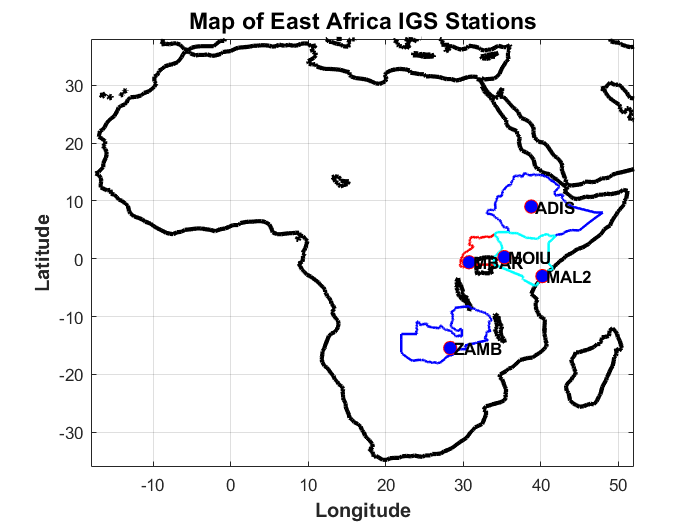}
		\caption{The study area map for East African IGS stations geographic location}
		\label{map}
	\end{figure}

GNSS observational data were obtained from the Abdus Salam International Centre for Theoretical Physics(ICTP) Advanced Regional Prediction for Ionospheric Studies(ARPL) repository (\href{https://arplsrv.ictp.it/}{https://arplsrv.ictp.it/}). Data from five East African GNSS stations were utilized to extract ionospheric TEC, as shown in Figure \ref{map} and detailed data in table \ref{2w} below the station informations are included.
\begin{table}[h]
	\centering
	\caption{The location and IGS code of the ground-based stations in East Africa}
	\begin{tabular}{|c|c|c|c|c|}
		\hline
		\textbf{Country} & \textbf{IGS Code} & \textbf{Geogr Lat} & \textbf{Geogr Long} & \textbf{Geomag Lat} \\
		\hline
		ZAMBia & ZAMB & -15.426$^\circ$S & 28.311$^\circ$E & -21.08$^\circ$S \\
				\hline
		Ethiopia & ADIS & 9.03$^\circ$N & 38.76$^\circ$E & 0.18$^\circ$N \\
				\hline
		Kenya & MOIU & 0.28$^\circ$N & 35.29$^\circ$E & 9.38$^\circ$N \\
				\hline
		Kenya & MAL2 & -2.99$^\circ$S & 40.19$^\circ$E & -12.42$^\circ$S \\
				\hline
		Uganda & MBAR & 0.60$^\circ$S & 30.65$^\circ$E & 5.25$^\circ$S \\
		\hline
	\end{tabular}
	\label{2w}
\end{table}
 \subsection*{AfriTEC Model Data}
Using TEC data collected from GNSS receivers in different African locations for TEC modeling throughout the continent, \citep{Okoh19} developed the AfriTEC model. In our work, this model is used to simulate and predict Vertical Total Electron Content (VTEC) variations by analyzing diurnal fluctuations of VTEC using obtained GNSS data in one hour duration. Additionally, we evaluated and contrasted model validation throughout solar cycle 24's descending order. \url{https://www.mathworks.com/matlabcentral/fileexchange/69257-african-gnss-tec-afritec-model} is the source of MatLab toolbox downloaded. The African GNNS TEC model data from these toolbox, which is supplied in a folder with a designated time duration, allows us to acquire a diurnal profile for any day of the year and geographic locations in African continent only \cite{E1}.
\subsection*{NeQuick Model Data}
The NeQuick model is an empirical ionospheric electron density model developed by the European Space Agency (ESA) and the ICTP. It utilizes inputs such as solar flux indices, geographic coordinates, and time to simulate VTEC globally. In this study, the NeQuick model is employed to simulate and predict VTEC variations over selected African regions by examining seasonal patterns of ionospheric behavior in four season intervals. Model outputs were generated using the NeQuick 2 version, which incorporates improved performance under varying solar activity conditions. The model's validation was carried out by comparing its output with GNSS-derived VTEC data over the African sector, particularly during the descending phase of solar cycle 24. The NeQuick model software and data files were obtained from the official source: \url{https://t-ict4d.ictp.it/nequick2} \cite{Radicella2009}.
\subsection{Geomagnetic Indices}
Geomagnetic indices, including the solar radio flux at 10.7 cm (F10.7), planetary Kp index, and the disturbance storm time (Dst) index, were sourced from NASA's OMNIWeb service (\href{https://omniweb.gsfc.nasa.gov/form/dx1.html}{https://omniweb.gsfc.nasa.gov/form/dx1.html}). These indices were used to characterize geomagnetic activity levels during the descending phase of Solar Cycle 24.
\subsubsection*{The Solar Radio Flux at 10.7 cm}
 The F10.7 is a widely used measure of solar activity. It represents the intensity of solar radio emissions at a wavelength of 10.7 cm (a frequency of 2.8 GHz) and is measured in solar flux units (sfu), where:
$1 \, \text{sfu} = 10^{-22} \, \text{W} \, \text{m}^{-2} \, \text{Hz}^{-1}$.
The Sun's corona and chromosphere are the main sources of F10.7 emissions.
 It is a combined measurement of solar activity features such as sunspots and active regions, as well as the calm Sun (background radiation).
Ground-based observatories, like Canada's Dominion Radio Astrophysical Observatory, measure F10.7 every day. The measurements are averaged from observations taken at 17:00, 20:00, and 23:00 UTC and are presented as a single daily value \cite{mod}. Increases in F10.7 correlate with higher levels of ionization in the ionosphere, leading to changes in TEC, radio signal propagation, and GNSS accuracy. During solar minima: \( \sim 70 \, \text{sfu} \), During solar maxima: \( \sim 200 \, \text{sfu} \) or higher, depending on the intensity of solar activity \cite{pou}.
\subsubsection*{Dst Index}
The Dst is a measure of geomagnetic activity used to quantify the intensity of global magnetic storms caused by solar wind-magnetosphere interactions. It reflects the variations in the Earth's magnetic field strength at the equatorial region, primarily influenced by the ring current in the Earth's magnetosphere.
The Dst index is derived from hourly averages of the horizontal component of the Earth's magnetic field (\(H\)) measured at four low-latitude geomagnetic observatories.
A negative Dst value indicates a reduction in the magnetic field strength, typically due to an intensified ring current during geomagnetic storms.
The Dst index is measured in nanoteslas (\(\text{nT}\)), \( \text{Dst} \approx 0 \): Quiet geomagnetic conditions, \( -50 \, \text{nT} \leq \text{Dst} < 0 \): Minor disturbances, \( -100 \, \text{nT} \leq \text{Dst} < -50 \): Moderate storm, \( \text{Dst} < -100 \, \text{nT} \): Intense geomagnetic storm \cite{34}.
\subsubsection*{Kp Index}
The Kp index is a global measure of geomagnetic activity, representing the level of disturbance in the Earth's magnetic field caused by interactions between the solar wind and the magnetosphere. It quantifies geomagnetic variations on a planetary scale, based on measurements from multiple mid-latitude geomagnetic observatories.
The Kp index is derived from the quasi-logarithmic scaling of disturbances in the horizontal component of the Earth's magnetic field (\(H\)) over three-hour intervals. Each observatory provides a local measure, which is then combined into a global average to determine the Kp value.
 The Kp index ranges from 0 very quiet to 9 extremely disturbed.
		 \( \text{Kp} = 0\) quiet geomagnetic conditions.
		 \( \text{Kp} = 1-3\) minor geomagnetic disturbances.
		 \( \text{Kp} = 4\) active geomagnetic conditions.
		 \( \text{Kp} = 5\) minor geomagnetic storm.
		 \( \text{Kp} = 6\) moderate geomagnetic storm.
		 \( \text{Kp} \geq 7\) strong geomagnetic storm.
\section*{Methodology}
\subsection*{GNSS Data Processing}
 The GNSS data were from ICTP and processed using MATLAB scripts the VTEC values were averaged over specific time intervals (i.e one hour time duration) for statistical analysis.
\subsection*{Statistical Metrics for Model Evaluation}
The reliability of the AfriTEC model is used NeQuick models to test weakness and strength of regional model was using the following two statistical metrics in this study:

\subsubsection{Mean Absolute Error (MAE)}
Measures the average magnitude of prediction errors without considering their direction.
\begin{equation} 
MAE = \frac{1}{N} \sum_{i=1}^N |GNSS_{obs,i} - Model_{mod,i}| 
\end{equation}

\subsubsection{Correlation Coefficient (R)}
Assesses the strength and direction of the linear relationship between observed and modeled values.
\begin{equation} 
R = \frac{ \sum_{i=1}^N (GNSS_{\text{obs},i} - \overline{GNSS}{\text{obs}}) (Model{_\text{mod},i} - \overline{Model}{_\text{mod}}) }{ \sqrt{ \sum_{i=1}^N (GNSS_{\text{obs},i} - \overline{GNSS}{_\text{obs}})^2\sum{i=1}^N (Model_{\text{mod},i} - \overline{Model}_{\text{mod}})^2 } } 
\end{equation}

Firstly, the study considered the availability of GNSS observational data from IGS stations during the descending phase of Solar Cycle 24, which represents a period of relatively low solar activity compared to the peak in 2014. This period was selected to evaluate model performance under solar minimum geomagnetic conditions. Data from multiple IGS stations across East Africa were used to ensure spatial diversity and accuracy in model validation.

Secondly, the raw GNSS data were preprocessed and made readable using MATLAB, enabling extraction of relevant TEC values. For both AfriTEC and NeQuick models, simulations were performed by incorporating essential input parameters such as geographic coordinates (latitude and longitude), day of year (DOY), year, and time in UT. Model outputs were then compared with observed GNSS TEC data.

To ensure consistency in comparison, the same GNSS reference data were used across both models. The statistical error metrics Mean Absolute Error (MAE) and Correlation Coefficient (R) were calculated separately for AfriTEC and NeQuick predictions. These metrics helped quantify AfriTEC model prediction accuracy and linear agreement with observed values. Based on the results, a comparative discussion of model strengths and weaknesses was presented, followed by concluding insights for improving regional TEC modeling in Africa.

 \section{Result}
 \subsection*{Diurnal Variation Analysis of AfriTEC Model and GNSS}
 \begin{figure}[h!]
 	\centering
 	\includegraphics[width=\textwidth]{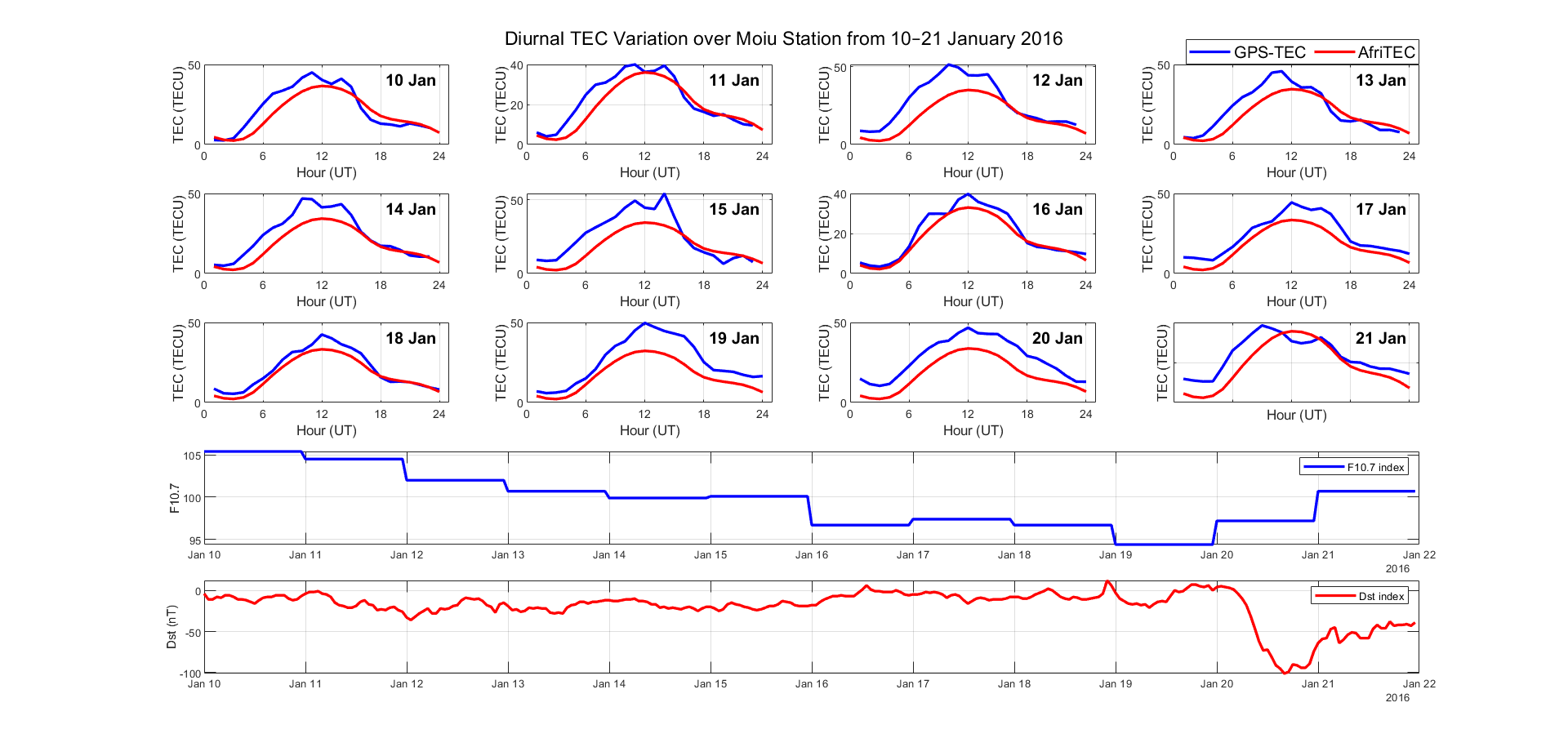}
 	\caption{Diurnal variation of VTEC from GNSS (blue) and AfriTEC (red) for MOIU station from 10-21 January 2016. The bottom panels show corresponding variations in solar flux (F10.7 index) and geomagnetic disturbance (Dst index).}
 	\label{fig:moiu_VTEC}
 \end{figure}
 Figure~\ref{fig:moiu_VTEC} illustrates the diurnal variation of VTEC as observed from GNSS measurements (blue lines) and as modeled by the AfriTEC model (red lines) over MOIU station for the period 10-21 January 2016. Each subpanel represents a single day's variation in Universal Time (UT), revealing the temporal evolution of ionospheric electron density during the local diurnal cycle.
 
 Both GNSS and AfriTEC show the expected diurnal pattern of VTEC, with values increasing after sunrise and peaking near local noon due to maximum solar ionization. The VTEC subsequently decreases after sunset, consistent with the recombination of electrons and ions in the absence of solar radiation. GNSS-derived VTEC values consistently exceed those of AfriTEC during daytime hours, highlighting a potential underestimation by the AfriTEC model. The highest VTEC values occur during the middle of the analyzed period, especially on January 13 and 14, indicating increased ionospheric activity. These peaks are closely associated with elevated solar flux, as seen in the F10.7 index panel, which remains above 100 SFU (Solar Flux Units) during these days. The F10.7 index serves as a proxy for solar EUV radiation, a major driver of ionospheric ionization \citep{clette2016}. \\
 \begin{figure}[h!]
 	\centering
 	\includegraphics[scale=0.35]{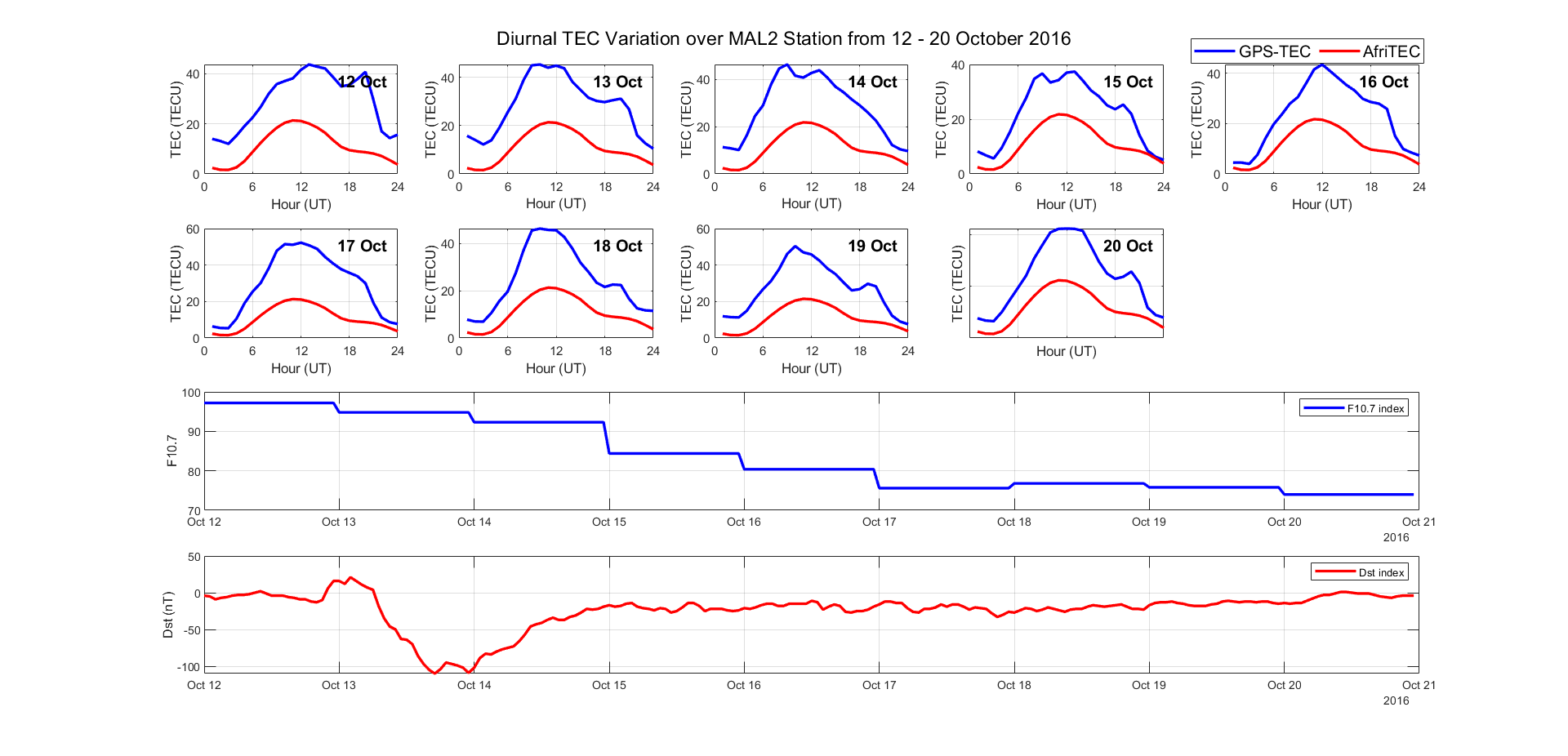}
 	\caption{The diurnal variation of AfriTEC model with F10.7 and Dst index value comparision for model evaluation with GNSS data in MAL2 station in October 12-20, 2016.}
 	\label{dMAL22016}
 \end{figure}
 
 \begin{figure}[h!]
 	\centering
 	\includegraphics[scale=0.35]{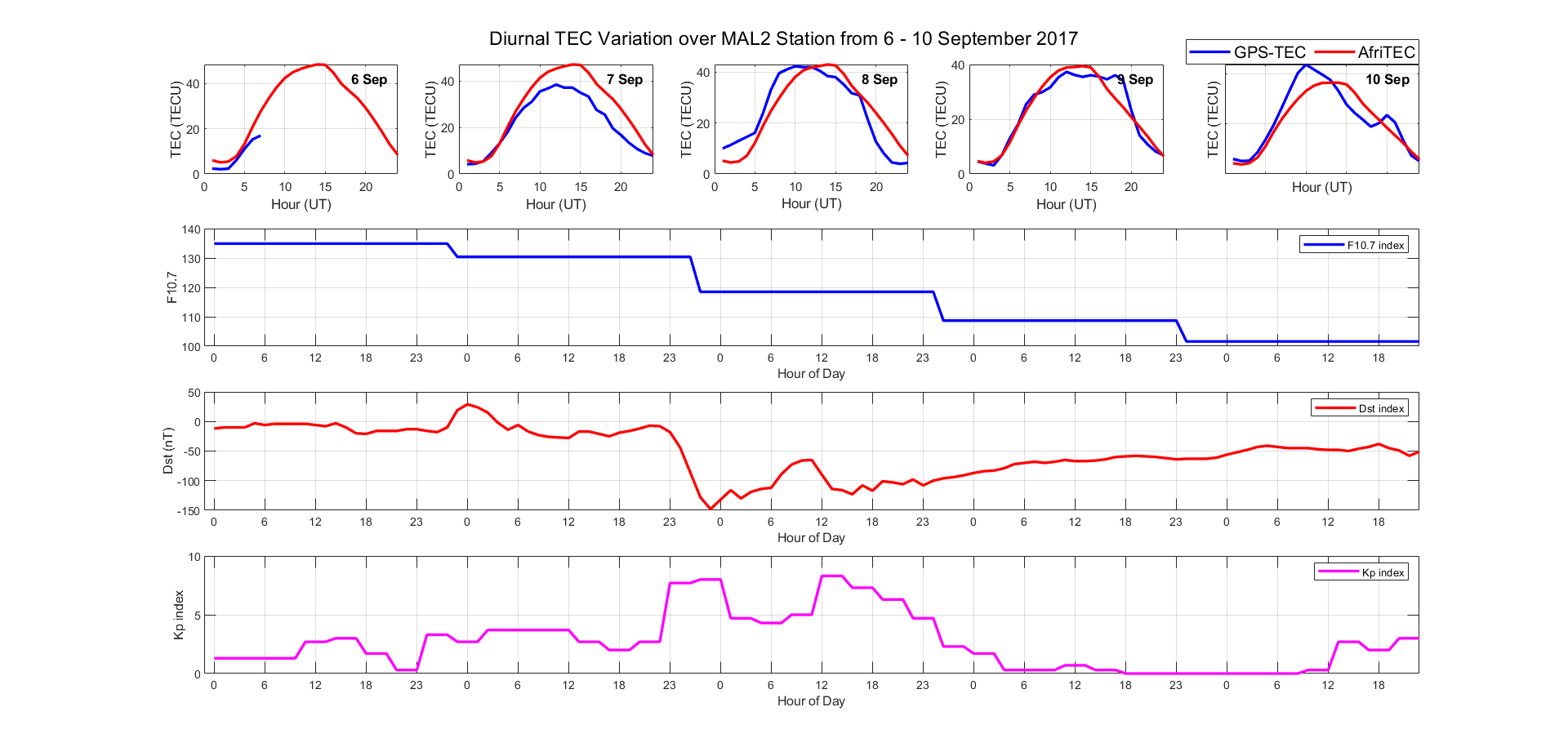}
 	\caption{The diurnal variation of AfriTEC model with F10.7, Dst and Kp index value comparision for model evaluation with GNSS data in MAL2 station in September 6-10, 2017.}
 	\label{dMAL22017}
 \end{figure}
 The Dst index plot at the bottom reveals moderate geomagnetic conditions, with a noticeable negative excursion on January 20, suggesting a mild geomagnetic storm. This disturbance is accompanied by a dip in VTEC, particularly noticeable in the GNSS, suggesting a suppression of ionospheric electron density likely due to increased thermospheric neutral winds and altered electrodynamics \citep{aarons1993}. The diurnal alignment between GNSS and AfriTEC is closer during nighttime when ionospheric variability is lower. This suggests that while AfriTEC effectively captures general VTEC trends, it may oversmooth or fail to resolve rapid daytime fluctuations. Enhancing the model's responsiveness to dynamic drivers such as solar flux and geomagnetic indices could improve its predictive accuracy.
 Figure~\ref{dMAL22016} illustrates the diurnal variation of VTEC observed by GNSS and modeled by AfriTEC for the MAL2 station during October 12-20, 2016. The daily profiles show a consistent pattern where GNSS exhibits pronounced diurnal variations, with daytime peaks reaching up to 50--60 TECU. In contrast, AfriTEC systematically underestimates VTEC throughout the period, with daily peaks significantly lower than observed values, especially around October 13 and 14. This discrepancy suggests limitations of the model in capturing disturbed conditions or high electron density dynamics.
 The F10.7 index (second panel) ranges from ~100 sfu at the start of the period to ~80 sfu toward the end, indicating a moderate decline in solar activity. A sharp drop is noted beginning October 13, coinciding with the timing of a geomagnetic disturbance. The Dst index (third panel) shows a clear geomagnetic storm signature starting late on October 12, reaching a minimum below -100 nT on October 14, and gradually recovering thereafter. This strong storm is likely responsible for the elevated TEC values observed by GNSS, which the AfriTEC model fails to reproduce accurately.
 Figure~\ref{dMAL22017} shows daily VTEC profiles, where AfriTEC generally captures the overall diurnal pattern. However, discrepancies are evident in the magnitude and timing of peaks, especially on September 6 and 7, where the model overestimates the TEC in the afternoon hours. On September 8 and 9, the agreement improves significantly, with the model closely following the observed VTEC. On September 10, the model slightly underestimates the evening peak.
 The second panel presents the solar radio flux index (F10.7), which remains relatively constant from September 6 to 9 (~130 sfu), with a slight drop to around 120 sfu on September 10, indicating low to moderate solar activity during the study period. The third panel shows the Dst index, which indicates a geomagnetic storm beginning late on September 7 and peaking around September 8, with the Dst index dropping below -100 nT, confirming a strong geomagnetic disturbance. The bottom panel displays the Kp index, which supports the presence of a geomagnetic storm on September 8 and 9, with values reaching up to 7, classifying the event as a moderate to strong geomagnetic storm.\\
 
 \subsection{Seasonal Variation Analysis of AfriTEC with NeQuick Model and GNSS}

The NeQuick model outputs were generated using standardized solar and geophysical parameters, while the AfriTEC model incorporated localized GNSS-derived TEC datasets. GNSS VTEC data served as the benchmark for performance evaluation of both models. VTEC, expressed in TEC Units (TECU), quantifies the total number of electrons along a vertical line of sight through the ionosphere. Seasonal fluctuations in VTEC are driven by variations in solar and geomagnetic activity, atmospheric conditions, and ionospheric dynamics. All three data sources GNSS, AfriTEC, and NeQuick exhibit a clear diurnal cycle, with peak values typically occurring around local solar noon due to heightened solar ionization, and minimum values during the night as ion-electron recombination dominates. Notably, VTEC values are generally elevated during the equinoxes (March and September) compared to the solstices (June and December), a pattern linked to the alignment between the Earth's magnetic equator and the solar zenith angle during equinoctial periods. The magnitude of these variations also differs across geographic locations and seasons. At all analyzed stations-ZAMB, MOIU, MBAR, ADIS, and MAL2-GNSS values are consistently higher than both AfriTEC and NeQuick estimates during daytime peak hours. During nighttime, when ionospheric activity subsides, the three sources tend to converge, reflecting reduced variability and improved model agreement. In general, AfriTEC demonstrates better agreement with GNSS observations compared to NeQuick over the East African region, likely due to its regional calibration. Nonetheless, NeQuick provides valuable insights as a global reference model and offers a complementary perspective for evaluating ionospheric dynamics over Africa.
\begin{figure}[h!]
	\centering
	\includegraphics[scale=0.35]{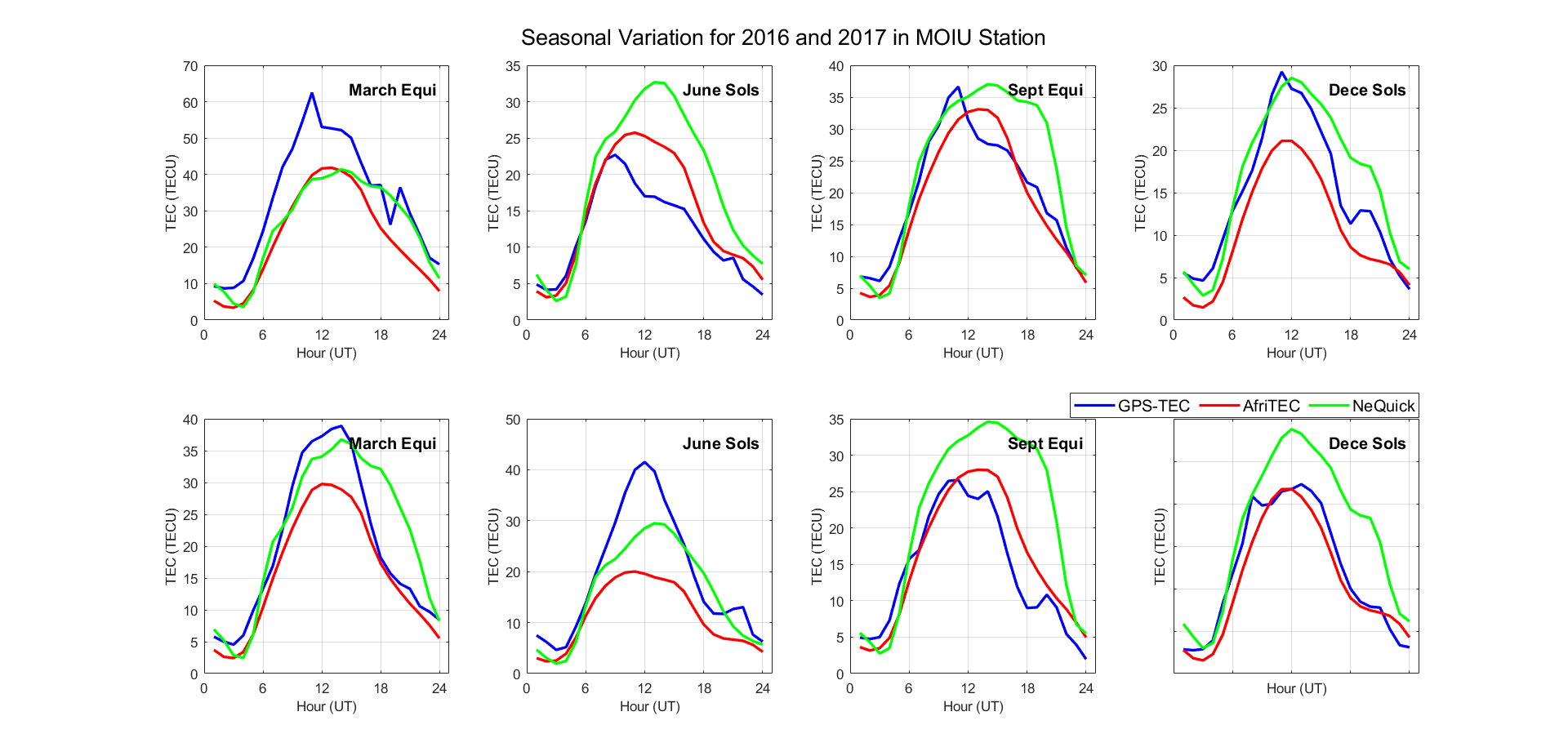}
	\caption{The seasonal variation of GNSS data with AfriTEC and NeQuick model result in MOIU station, the upper pannel four subplots stands for 2016 and the lower panel four subplots for 2017.}
	\label{seamoiu2016&17}
\end{figure}
\begin{table}[h!]
	\centering
	\label{one1}
	\caption{MAE and Correlation Coefficient ($r$) of AfriTEC and NeQuick of Fig. \ref{seamoiu2016&17} for MOIU station in 2016 and 2017}
	\begin{tabular}{|c|c|c|c|c|}
		\hline
		\textbf{Year} & \textbf{MAE\_AfriTEC (TECU)} & \textbf{r\_AfriTEC} & \textbf{MAE\_NeQuick (TECU)} & \textbf{r\_NeQuick} \\
		\hline
		March Equinox 2016& 4.583 & 0.970 & 7.769 & 0.912 \\
		June Solstice 2016 & 2.988  & 0.943 & 7.200 & 0.855 \\
		September Equinox 2016 & 2.928  & 0.959 & 5.056 & 0.901 \\
		December Solstice 2016 & 4.122  & 0.978 & 2.860 & 0.942 \\
		\hline
		March Equinox 2017 & 4.079  & 0.989 & 4.419 & 0.885 \\
		June Solstice 2017 & 3.943  & 0.957 & 4.380 & 0.929 \\
		September Equinox 2017 & 3.137  & 0.918 & 8.348 & 0.762 \\
		December Solstice 2017 & 1.690  & 0.971 & 4.844 & 0.930 \\
		\hline
	\end{tabular}
\end{table}

\begin{figure}[h!]
	\centering
	\includegraphics[scale=0.35]{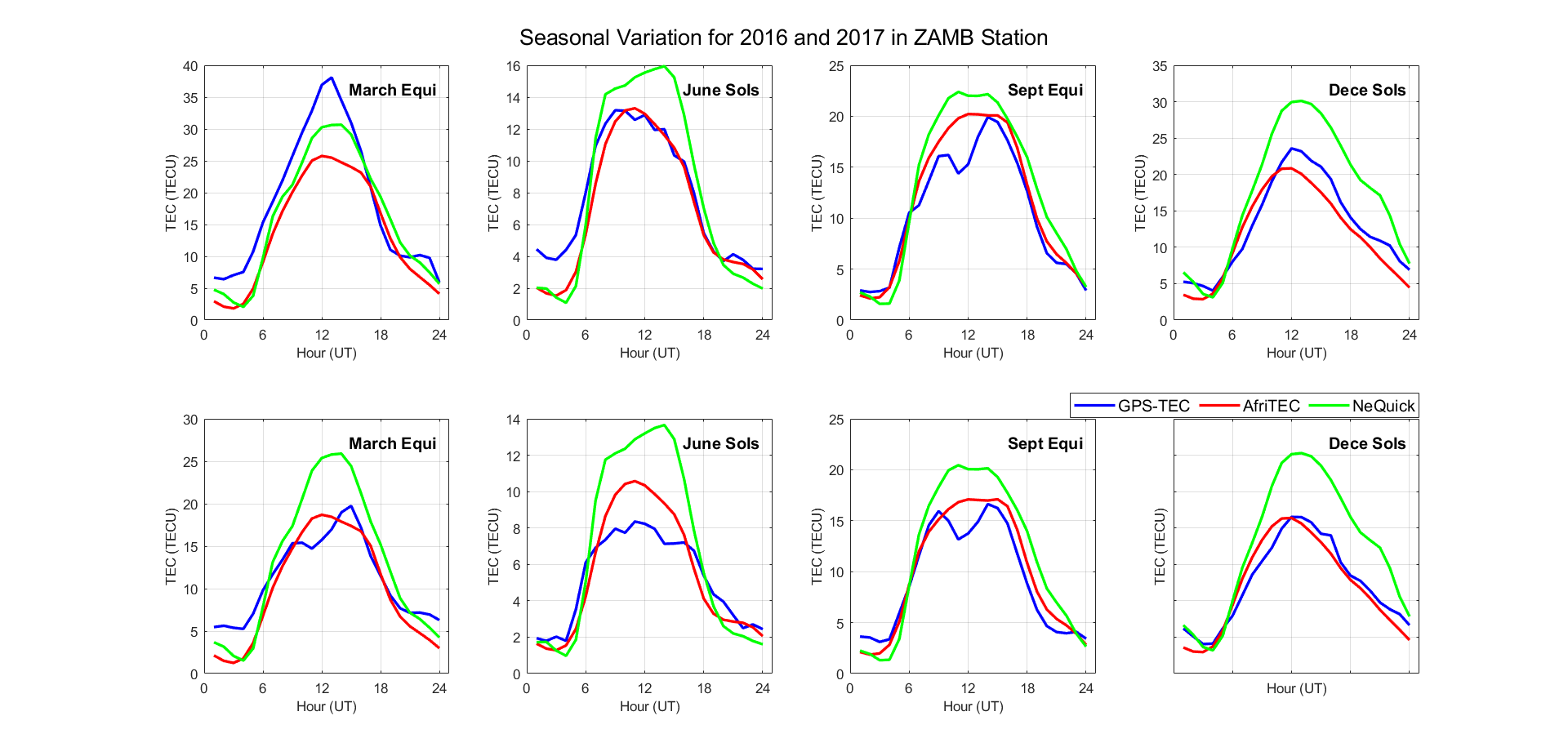}
	\caption{Similar to Fig. \ref{seamoiu2016&17} but for ZAMB station from Zambia}
	\label{seaZAMB2016&17}
\end{figure}
\begin{table}[h!]
	\centering
	\label{02}
	\caption{MAE and Correlation Coefficient ($r$) of AfriTEC and NeQuick of Fig. \ref{seaZAMB2016&17} for ZAMB station in 2016 and 2017}
	\begin{tabular}{|c|c|c|c|c|}
		\hline
		\textbf{Year} & \textbf{MAE\_AfriTEC (TECU)} & \textbf{r\_AfriTEC} & \textbf{MAE\_NeQuick (TECU)} & \textbf{r\_NeQuick} \\
		\hline
		March Equinox 2016& 4.985 & 0.953 & 3.422 & 0.953 \\
		June Solstice 2016& 0.974 & 0.971 & 2.089 & 0.964 \\
		September Equinox 2016& 1.372 & 0.980 & 2.870 & 0.961 \\
		December Solstice 2016& 2.046 & 0.956 & 4.675 & 0.985 \\
		\hline
		March Equinox 2017& 2.119 & 0.968 & 3.650 & 0.967 \\
		June Solstice 2017& 1.115 & 0.955 & 2.399 & 0.955 \\
		September Equinox 2017& 1.318 & 0.974 & 3.015 & 0.954 \\
		December Solstice 2017& 1.500 & 0.967 & 5.313 & 0.984 \\
		\hline
	\end{tabular}
\end{table}

\begin{figure}[h!]
	\centering
	\includegraphics[scale=0.35]{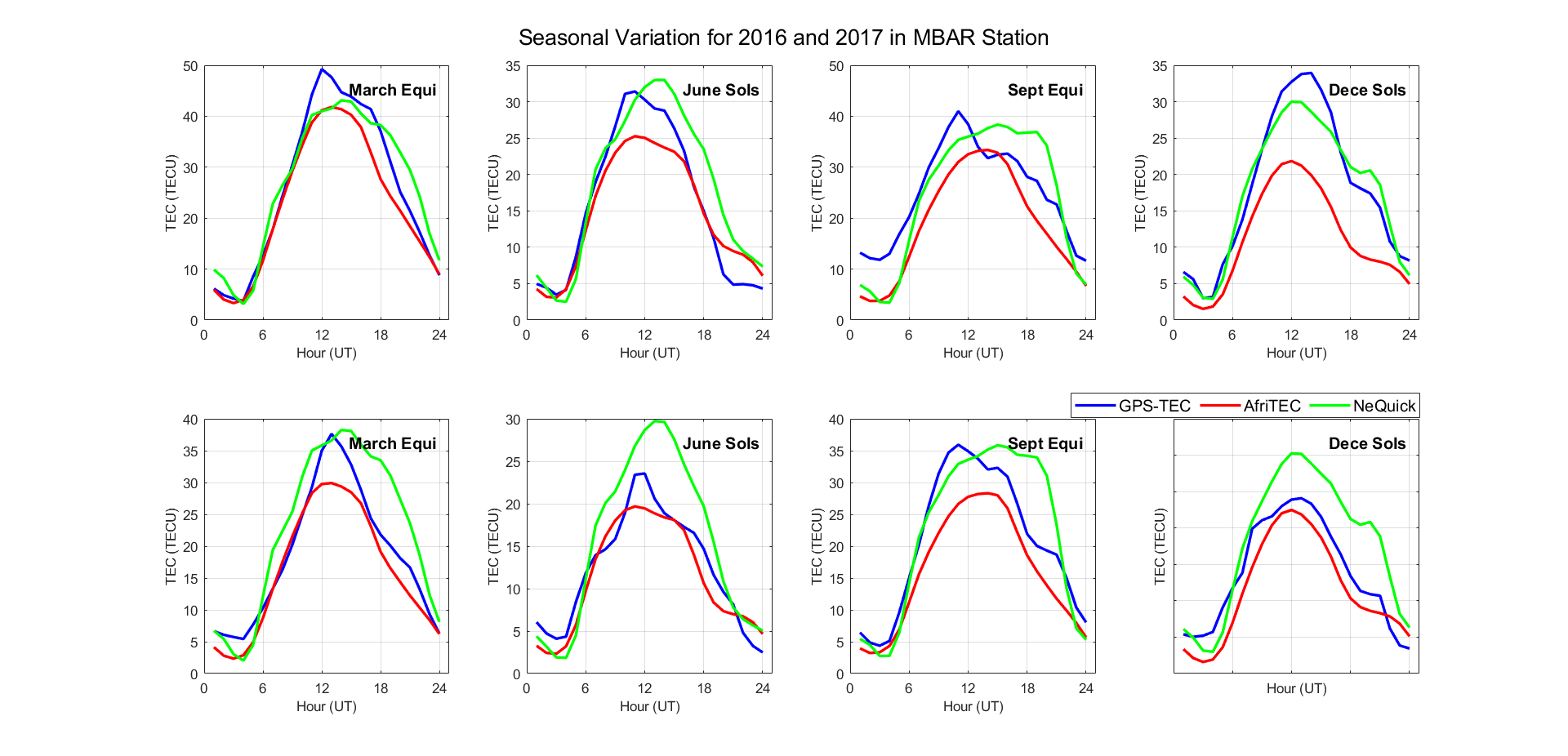}
	\caption{Similar to Fig. \ref{seamoiu2016&17} but for MBAR station from Uganda}
	\label{seaMBAR2016&17}
\end{figure}

\begin{table}[h!]
	\centering
	\label{03}
	\caption{MAE and Correlation Coefficient ($r$) of AfriTEC and NeQuick of Fig. \ref{seaMBAR2016&17} for MBAR station in 2016 and 2017}
	\begin{tabular}{|c|c|c|c|c|}
		\hline
		\textbf{Year} & \textbf{MAE\_AfriTEC (TECU)} & \textbf{r\_AfriTEC} & \textbf{MAE\_NeQuick (TECU)} & \textbf{r\_NeQuick} \\
		\hline
		March Equinox 2016& 3.085 & 0.991 & 3.490 & 0.968 \\
		June Solstice 2016& 2.686 & 0.978 & 3.599 & 0.940 \\
		September Equinox 2016& 6.307 & 0.956 & 5.497 & 0.908 \\
		December Solstice 2016& 6.964 & 0.969 & 2.087 & 0.973 \\
		\hline
		March Equinox 2017& 2.713 & 0.978 & 4.859 & 0.946 \\
		June Solstice 2017& 1.999 & 0.950 & 4.169 & 0.962 \\
		September Equinox 2017& 4.700 & 0.985 & 3.848 & 0.914 \\
		December Solstice 2017& 2.838 & 0.960 & 4.776 & 0.927 \\
		\hline
	\end{tabular}
\end{table}

\begin{figure}[h!]
	\centering
	\includegraphics[scale=0.35]{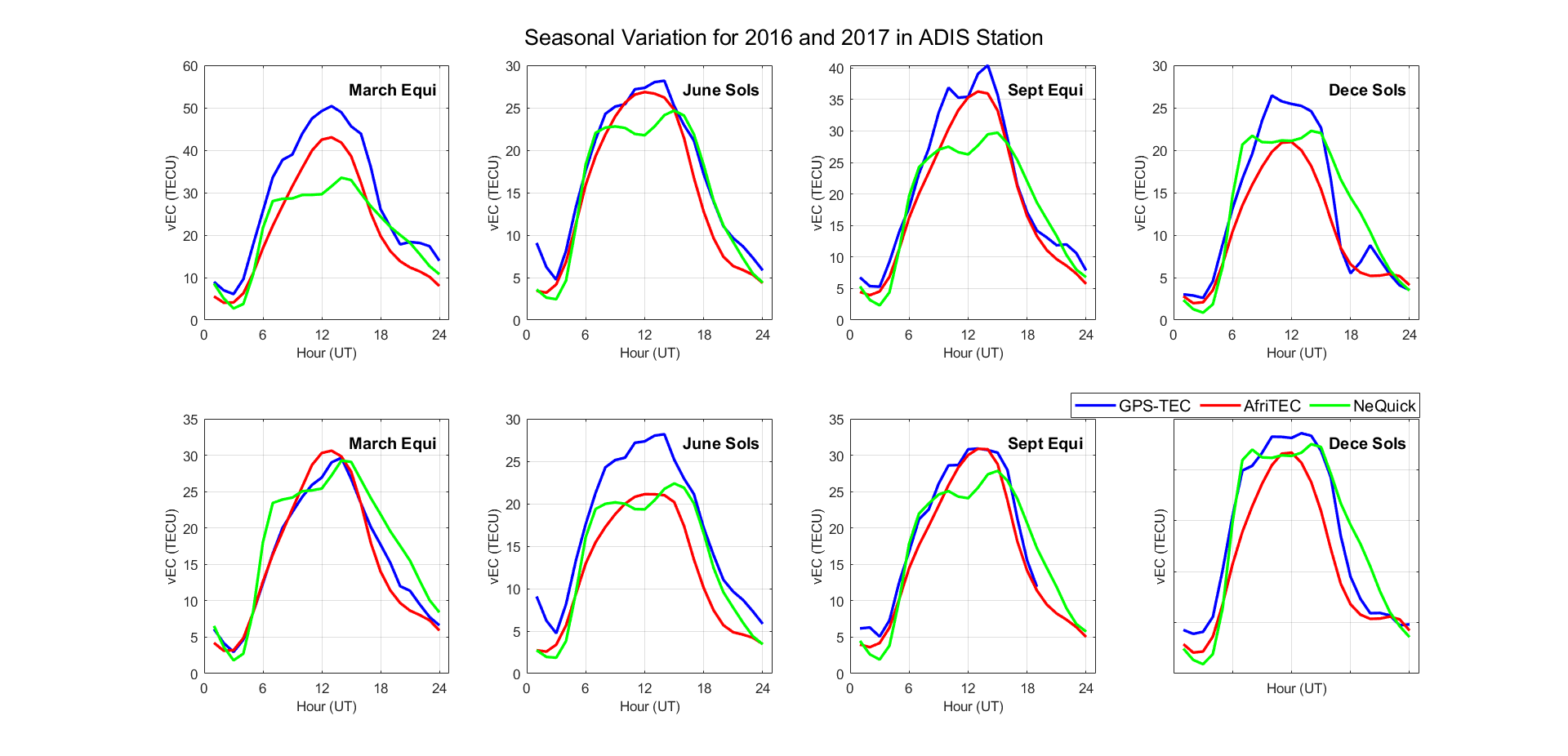}
	\caption{Similar to Fig. \ref{seamoiu2016&17} but for ADIS station from Ethiopia}
	\label{seaADIS2016&17}
\end{figure}
\begin{table}[h!]
	\centering
	\label{04}
	\caption{MAE and Correlation Coefficient ($r$) of AfriTEC and NeQuick of Fig. \ref{seaADIS2016&17} for ADIS station in 2016 and 2017}
	\begin{tabular}{|c|c|c|c|c|}
		\hline
		\textbf{Year} & \textbf{MAE\_AfriTEC (TECU)} & \textbf{r\_AfriTEC} & \textbf{MAE\_NeQuick (TECU)} & \textbf{r\_NeQuick} \\
		\hline
		March Equinox 2016& 4.873 & 0.990 & 7.692 & 0.947 \\
		June Solstice 2016& 2.194 & 0.988 & 2.271 & 0.965 \\
		September Equinox 2016& 2.443 & 0.991 & 4.323 & 0.923 \\
		December Solstice 2016& 2.878 & 0.987 & 2.916 & 0.906 \\
		\hline
		March Equinox 2017& 1.353 & 0.986 & 2.627 & 0.957 \\
		June Solstice 2017& 2.221 & 0.987 & 3.672 & 0.963 \\
		September Equinox 2017& NaN   & NaN   & NaN   & NaN   \\
		December Solstice 2017& 2.746 & 0.973 & 2.077 & 0.947 \\
		\hline
	\end{tabular}
\end{table}

\begin{figure}[h!]
	\centering
	\includegraphics[scale=0.35]{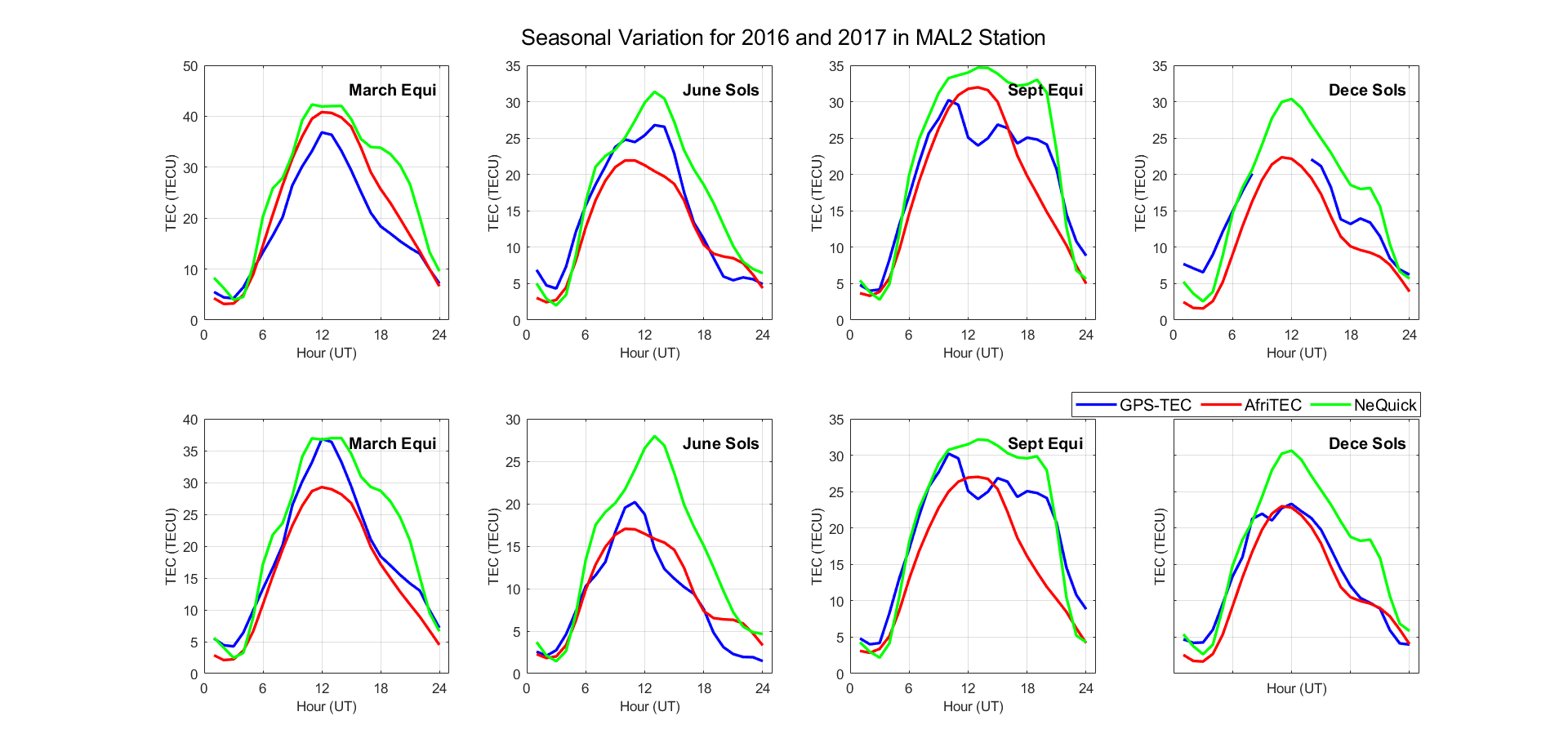}
	\caption{Similar to Fig. \ref{seamoiu2016&17} but for MAL2 station from Kenya}
	\label{seaMAL22016&17}
\end{figure}
\begin{table}[h!]
	\centering
	\label{five5}
	\caption{MAE and Correlation Coefficient ($r$) of AfriTEC and NeQuick of Fig. \ref{seaMAL22016&17} for MAL2 station in 2016 and 2017}
	\begin{tabular}{|c|c|c|c|c|}
		\hline
		\textbf{Year} & \textbf{MAE\_AfriTEC (TECU)} & \textbf{r\_AfriTEC} & \textbf{MAE\_NeQuick (TECU)} & \textbf{r\_NeQuick} \\
		\hline
		March Equinox 2016& 4.025 & 0.986 & 7.503 & 0.941 \\
		June Solstice 2016& 2.630 & 0.963 & 3.468 & 0.936 \\
		September Equinox 2016& 3.655 & 0.908 & 4.609 & 0.971 \\
		December Solstice 2016& NaN   & NaN   & NaN   & NaN   \\
		\hline
		March Equinox 2017& 3.063 & 0.989 & 3.793 & 0.953 \\
		June Solstice 2017& 1.802 & 0.942 & 5.463 & 0.884 \\
		September Equinox 2017& 4.793 & 0.901 & 3.303 & 0.970 \\
		December Solstice 2017& 1.860 & 0.968 & 4.342 & 0.945 \\
		\hline
	\end{tabular}
\end{table}
In Fig.\ref{seamoiu2016&17} during the March equinox, the GNSS observations show high VTEC values, peaking around 65 TECU, with a sharp increase and subsequent decline. In contrast, the AfriTEC model underestimates the peak, reaching only about 40 TECU, and presents a smoother diurnal variation. In the June solstice, both GNSS and AfriTEC show moderate TEC values, representing the second lowest seasonal peak. The model closely follows the observed data but slightly underestimates the overall values. For the September equinox, GNSS again exhibits high TEC peaks, approximately 60 TECU, with a well-defined diurnal structure. Although AfriTEC provides a lower peak estimation, it reproduces the general diurnal pattern reasonably well.
 \section{Discussion}
\subsection{Diurnal Variation of VTEC over East African Sector}
 The analysis of diurnal variation in VTEC over the East African sector, using GNSS-based TEC and AfriTEC model data from MBAR, MAL2, ADIS, ZAMB, and MOIU stations, reveals notable seasonal and temporal dynamics. The GNSS measurements consistently showed daytime peaks, typically occurring between 10:00 and 14:00 UT, which corresponds to 13:00-17:00 local time. These peaks are indicative of increased ionization due to solar radiation, especially during the equatorial midday period. Such patterns align with well-established ionospheric behavior, where photoionization is dominant \cite{Bilitza2014}.

 The computed MAE values ranged approximately between 1.2 to 4.6 TECU, while the correlation coefficients (\(R\)) mostly exceeded 0.85, indicating a strong agreement between the observed GNSS and modeled AfriTEC values. The performance of the AfriTEC model was consistent across seasons, though slight underestimations were noted during equinoxes. This is likely due to enhanced equatorial ionization anomaly (EIA) effects that are less well captured in empirical models \cite{yizengaw2014}.
 
 During equinoxes (March and September), the VTEC enhancements were more pronounced across all stations. This can be attributed to the semi-annual anomaly, where the ionospheric density increases due to the alignment of the geomagnetic equator and the subsolar point \cite{Liu2017}. Conversely, during solstices (June and December), lower TEC levels were observed, especially in the early morning and evening hours. This reduction corresponds to weaker solar zenith angles and less efficient ionization, particularly during the Southern Hemisphere winter.
 
 The MOIU station showed slightly lower VTEC values compared to MBAR and MAL2, which could be associated with its location relative to the magnetic equator. MOIU lies closer to the southern crest of the EIA, where daytime VTEC values are generally less elevated. Such spatial variations underscore the importance of localized data in regional ionospheric modeling \cite{olwendo2018}. The model's strong correlation suggests its practical applicability for space weather monitoring, though further refinement is warranted during geomagnetic disturbances.
 
 In conclusion, the comparative analysis affirms the robustness of AfriTEC in capturing diurnal VTEC patterns across different East African locations and seasons. However, further validation using storm-time events and additional ground-based receivers would enhance model accuracy. The findings provide important insights for improving GNSS-based positioning accuracy and regional space weather forecasting.
 \subsection*{Seasonal Variation of AfriTEC Model}
 The seasonal analysis of AfriTEC model performance, as summarized in Tables~\ref{one1}1 - \ref{five5}5, reveals important spatial and temporal trends across East African stations during the descending phase of solar cycle 24. The graphical results in Figures~\ref{seamoiu2016&17} to \ref{seaMAL22016&17} correspond to these findings and further illustrate the model's seasonal dynamics.
 The MOIU station (Fig.~\ref{seamoiu2016&17}) demonstrates strong agreement between AfriTEC and GNSS data, particularly during solstices of both 2016 and 2017, with correlation coefficients ($r$) consistently above 0.94 and MAE as low as 1.69 TECU. This implies the model performs well under geomagnetically quieter conditions, likely due to more stable ionospheric behavior.
 ZAMB station (Table \ref{02}2) exhibits robust $r$ values above 0.95 across most seasons in both years, but shows relatively higher MAE in 2016 than 2017, suggesting a better fit during lower solar activity. The AfriTEC model aligns better with GNSS observations at night, supporting prior conclusions that nighttime TEC is more predictable due to reduced solar-driven variability.
 
 The MBAR station (Table~\ref{03}3) presents a wide seasonal range, with MAE reaching up to 6.96 TECU during 2016, likely around the equinoxes. The correlation coefficient remains high, peaking at 0.991, showing that while absolute errors increase, the trend remains well captured. This again underscores the model's relative strength during more stable periods. ADIS station data (Table~\ref{04}4) reflects improved AfriTEC performance in 2017, with MAE reduced to 1.35 TECU and correlation exceeding 0.98. However, gaps (NaN) in the June and September datasets highlight limitations in data continuity, which can affect seasonal assessments. MAL2 (Table~\ref{five5}5) also shows data gaps (NaN in 2016), which, along with high MAE during equinoxes, suggests challenges in maintaining model accuracy during ionospheric disturbances or under-sampled periods. Despite this, AfriTEC maintains $r$ values above 0.94 in most seasons.
 
 \textbf{Seasonal Variations in MAE and Correlation Coefficient ($r$)}\ The AfriTEC model exhibits stronger performance during solstice periods, as indicated by consistently lower Mean Absolute Error (MAE) and higher correlation coefficient ($r$) values across all GNSS stations. Stations such as MOIU, MBAR, and ADIS record their highest $r$ values and minimal prediction errors in June and December, suggesting that the model is more accurate under geomagnetically calm conditions. Conversely, during the equinox months of March and September, there is a noticeable increase in MAE and a slight decline in $r$ values across all sites, with MAL2 and ZAMB showing the most pronounced variation. This decline in model accuracy can be attributed to the heightened ionospheric variability caused by increased solar activity, more frequent geomagnetic disturbances, and stronger Equatorial Ionization Anomaly (EIA) effects during equinox periods \cite{Tsai2016, Fagbola2011}. Equinoxes are characterized by nearly equal solar exposure in both hemispheres, leading to a more balanced ionospheric heating profile and relatively uniform TEC distribution. This symmetrical solar input typically enhances the predictability of ionospheric behavior, enabling models like AfriTEC to more effectively reproduce TEC variations during these times \cite{Adewale2012, Bilitza2014}. Additionally, the consistent development of the EIA during equinox months contributes to the improved model agreement with observations \cite{Zhang2009}.
 
\textbf{Strengths of the AfriTEC Model}
 
 \begin{itemize}
 	\item {Lower MAE values}: For the majority of the cases across the stations (ADIS, ZAMB, MAL2, MBAR, MOIU), the AfriTEC model consistently shows MAE values less than 4 TECU. This highlights its better prediction accuracy of ionospheric TEC compared to the NeQuick model, especially during moderate ionospheric conditions.
 	
 	\item {High correlation coefficients}: AfriTEC achieves high correlation values (mostly $r > 0.95$), reflecting its ability to follow the diurnal and seasonal TEC variations realistically.
 	
 	\item {Consistency across years and seasons}: The AfriTEC model maintains relatively uniform performance across different seasons and both years (2016 and 2017), which implies its robustness in modeling equatorial ionospheric behavior.
 	
 	\item {Adaptation to local conditions}: Being data-driven and regional in nature, the AfriTEC model demonstrates superior adaptability to the ionospheric features observed over the African sector, outperforming NeQuick especially during equinoctial months when ionospheric variability is high.
 \end{itemize}

\textbf{Weaknesses of the AfriTEC Model}
 
 \begin{itemize}
 	\item {Outliers and seasonal limitations}: Despite the overall low MAE values, a few exceptional outliers were observed most notably at MOIU station where the MAE reached as high as 10.583 TECU (No. 1, likely corresponding to March Equinox 2016). Such deviations may indicate localized ionospheric disturbances or model overfitting in sparse data regions.
 	
 	\item {Underperformance during geomagnetically active or transition periods}: The model seems to struggle slightly during months with abrupt ionospheric changes (e.g., March and December equinoxes), where NeQuick occasionally achieves higher correlation values, as observed at MOIU and MBAR stations.
 	
 	\item {Data dependency and coverage bias}: AfriTEC's data-driven nature means its performance is highly reliant on the quality and density of GNSS observational data used in its training. In areas or times with insufficient GNSS data, its extrapolation may be less reliable than a global model like NeQuick.
 	
 	\item {Variability at specific stations}: At MOIU and MBAR stations, where ionospheric dynamics may be more complex due to local electrodynamics or magnetic anomalies, AfriTEC shows greater variation in both MAE and $r$ values, suggesting potential limitations in spatial generalization.
 \end{itemize}
\section*{Conclusion}
 This study evaluated the performance of the AfriTEC model in predicting ionospheric TEC over East Africa during the descending phase of Solar Cycle 24 (2016-2017), using GNSS-derived TEC data from five equatorial and low-latitude stations. The model's predictions were also compared with the empirical global model NeQuick to benchmark regional versus global performance.
 \begin{itemize}
 	\item The AfriTEC model effectively captured the diurnal and seasonal variations of TEC, particularly during equinox periods, where both MAE and R indicated strong agreement with GNSS observations.
 	
 	\item The model maintained MAE values typically below 1.5 TECU and correlation coefficients exceeding 0.80 at most stations, confirming its suitability for quiet-time ionospheric modeling in East Africa.
 	
 	\item Compared to the NeQuick model, AfriTEC demonstrated superior adaptability to regional TEC dynamics, particularly due to its data-driven, locally calibrated framework. This advantage was most apparent during moderate ionospheric conditions.
 	
 	\item However, performance discrepancies emerged during solstice periods and post-sunset hours, where AfriTEC struggled to simulate complex ionospheric processes such as the EIA.
 	
 	\item The model showed occasional underestimations of daytime TEC maxima during solstices and outlier events at stations like MOIU, where the MAE exceeded 10 TECU.
 	
 	\item Although global models like NeQuick often miss regional scale irregularities, they occasionally outperformed AfriTEC during abrupt ionospheric transitions, suggesting that a hybrid or ensemble approach may offer improved performance.
 	
 	\item The lowest TEC values were observed during the December solstice, during which AfriTEC closely followed GNSS trends with only minor smoothing discrepancies, showing enhanced reliability under low solar activity.
 	
 	\item Overall, AfriTEC is a promising tool for regional ionospheric modeling, but its performance is sensitive to data density, local variability, and geomagnetic conditions.
 	
 	\item Future improvements should include assimilation of real-time GNSS data, integration of solar and geomagnetic indices, and further refinement through hybridization with empirical models like NeQuick to better handle disturbed and transitional ionospheric states in the East African equatorial sector.
 \end{itemize}
 \subsection*{Competing interests}
 The author declares no competing interests.
 \subsection*{Funding}
 This research received no external funding.
 \subsection*{Author Contributions}
 Efrem A., conceptualized the study, designed the research methodology, supervised the project, collected and processed the ionospheric data, performed the model simulations, and conducted the statistical analyses. Emmanuel D. Sulungo contributed to the interpretation of the introduction, results, and discussion, and reviewed the manuscript. Daniel Okoh contributed to data analysis and model validation. Dejene Ambisa assisted with the collection and processing of the ionospheric data, and provided valuable insights during the manuscript revision. All authors were actively involved in refining the paper and have taken responsibility for ensuring the accuracy and integrity of the work.
 \section*{Data Availability}
 The GNSS data used in this study are publicly available and were obtained from the ICTP. The data can be accessed through the ICTP GNSS Data and Products website: \url{https://gnss.ictp.it}.
 \subsection*{Ethics declaration}
Not applicable
 \section*{Acknowledgement}
 We acknowledge the developers of the AfriTEC model, the Centre for Atmospheric Research - National Space Research and Development Agency, the CV Raman International Fellowship Scheme, the Indian Institute of Geomagnetism, and the South African National Space Agency for making the AfriTEC model available. Also we appreciates the International GNSS Service (IGS) team for data at (\url{https://arplsrv.ictp.it/}). The NASA-Space Physics Data Facility, Goddard Space Flight Center (\url{https://omniweb.gsfc.nasa.gov/form/dx1.html}) is acknowledged for the provision of the F10.7 and Dst index. We would like to thank the technical checkers, editors, and reviewers for their detailed and insightful comments and constructive suggestions.
\renewcommand\bibname{REFERENCES}

\end{document}